\documentclass[3p,twocolumn,times,sort&compress]{elsarticle}

\usepackage[utf8]{inputenc}
\usepackage{amsmath}
\usepackage{xspace}
\usepackage{graphicx}
\usepackage{textcomp}
\usepackage{url}

\graphicspath{{Figures/}}
\newcommand{\Rsq}{R$^2$\xspace}
\newcommand{\Eg}{$E_\text{g}$\xspace}
\newcommand{\Egp}{$E_\text{g}'$\xspace}
\newcommand{\del}{$\Delta$\xspace}
\newcommand{\eps}{$\varepsilon$\xspace}
\newcommand{\x}{$\xi$\xspace}

\begin{document}

%%%%%%%%%%

\title{Automated approaches for band gap mapping in STEM-EELS}
\author{Cecilie S. Granerød\corref{cor1}}
\ead{cecilie.granerod@fys.uio.no}
\author{Wei Zhan\corref{cor2}}
\author{Øystein Prytz\corref{cor2}}
\address{Department of Physics, Centre for Materials Science and Nanotechnology,\\University of Oslo, P. O. Box 1048 Blindern, N-0316 Oslo, Norway}
\cortext[cor1]{Corresponding author.}
\date{\today}

%%%%%

\begin{abstract}
Band gap variations in thin film structures, across grain boundaries, and in embedded nanoparticles are of increasing interest in the materials science community. As many common experimental techniques for measuring band gaps do not have the spatial resolution needed to observe these variations directly, probe-corrected Scanning Transmission Electron Microscope (STEM) with monochromated Electron Energy-Loss Spectroscopy (EELS) is a promising method for studying band gaps of such features. However, extraction of band gaps from EELS data sets usually requires heavy user involvement, and makes the analysis of large data sets challenging. Here we develop and present methods for automated extraction of band gap maps from large STEM-EELS data sets with high spatial resolution while preserving high accuracy and precision.
\end{abstract}

%%%%%

\begin{keyword}
STEM \sep EELS \sep Spectrum Images \sep Band gap measurements 
\end{keyword}

%%%%%

\maketitle

%%%%%%%%%%

\section{Introduction}	

The optical band gap, defined as the onset of absorption in a semiconductor material, is a central property in the development and improvement of a large number of technologies. Many common techniques for measuring band gaps in semiconductors are based on interaction with light and hence do not offer a spatial resolution better than on a \textmu m scale\cite{visser_high_1990, fukura_high_2005, castellanos-gomez_spatially_2016, ong_high-spatial-resolution_2000}. However, the demand for high efficiency and small size of new devices requires a fundamental understanding of the material and band gaps at smaller length scales. With a (Scanning) Transmission Electron Microscopes ((S)TEM's), band gap measurements performed using Electron Energy-Loss Spectroscopy (EELS) have a spatial resolution which in principle is limited only by the delocalization of the energy transfer process\cite{erni_valence_2005, egerton_limits_2007, bosman_nanoscale_2009, zhan_nanoscale_2017_fix}. For band gaps in common semiconductors, this delocalization is on the order of 5-10 nm\cite{egerton_limits_2007}, thereby making band gap measurements in TEM a powerful method for studying new semiconductor devices. 

In low-loss EELS the energy lost by the transmitted electron corresponds to energy transfer for excitations in the sample. The dominating feature in EELS is usually the zero-loss peak (ZLP), which contains transmitted electrons that have lost little or no energy to the specimen. The tail of this feature forms a background to the excitations which one usually wishes to study. If only single electron transitions from the valence to the conduction band are considered, the energy loss is related to the joint density of states, and the minimum of the observed energy loss corresponds to the optical band gap of the specimen. 

The task of measuring the band gap using EELS is then to identify the onset of energy loss in a precise and accurate manner, a process that usually relies on fitting of models for both the background and edge to the experimentally obtained data. In performing the fitting, several choices must be made and complications may arise: 
\begin{enumerate}[a)]
\item To achieve the highest accuracy in band gap extraction, a good model for the background (ZLP) should ideally be fitted as close to the edge onset as possible\cite{keller_local_2014}. If the onset shifts, so should the background fitting region.
\item An optimal energy range must be identified for fitting of the energy loss model to the data. As with the background modelling, this energy range will vary with the position of the edge onset (the band gap value)\cite{bosman_nanoscale_2009, bosman_mapping_2006}. Depending on the shape of the edge, this fitting region may also be relatively small, making precise positioning of the fit region important. 
\item Even after background subtraction, intensity may remain below the onset of the band gap transitions. This may be due to losses to Cherenkov radiation, excitations of surface plasmons, guided light modes, amorphous surface effects, or transitions to and from defect states\cite{lazar_materials_2003, gu_band-gap_2007, park_bandgap_2009}. In any band gap extraction process a determination needs to be done on how to handle such effects.
\item Depending on the experimental conditions, the obtained spectra may have a high level of noise, both in the edge itself and in the energy loss range below\cite{keller_local_2014, kimoto_advantages_2005, keast_new_2007}. The noise level must be evaluated and factored into the errors reported together with the extracted band gap values.
\end{enumerate}

If only a small data set containing a handful of spectra is analysed, each spectrum can be manually inspected and an optimal band gap extraction strategy can be identified individually. However, if the data is collected in a 2D scan with a spectrum at each scanned position, called Spectrum Image mode, the data set can contain thousands of spectra. The manual approach is then not viable. Furthermore, if the spectra are obtained from different parts of a chemically inhomogeneous specimen, band gap extractions based on one or a few optimal parameters are likely to give inaccurate results.

STEM-EELS has previously been applied in measuring band gaps with high spatial and energy resolution\cite{bosman_nanoscale_2009, gu_mapping_2009}, but very few works have attempted to put forward an efficient computational method for band gap mapping\cite{zhan_nanoscale_2017_fix}. In this work, an automated approach for STEM-EELS band gap analysis is presented, with the aim of extracting band gap maps from large Spectrum Images. Methods, referred to as the Sliding interval method, the Fixed endpoint method, and the Dynamic background subtraction method, have been implemented in MATLAB and are available on Github\cite{cecilsgr/eels-mapping-program}.

%%%%%%%%%%

\section{Experimental methods}

ZnO has a direct optical band gap of 3.25-3.30~eV at room temperature, and Cd-alloying has been shown to increase the lattice spacing, thereby decreasing the band gap\cite{wang_band_2006, zhu_electronic_2008, detert_crystal_2013_fix}. An MOVPE-grown sample with films of ZnO and Zn$_{1-x}$Cd$_x$O ($x\simeq0.1-0.2$, referred to as ZnCdO) was prepared by cutting and mechanical polishing, before ion milling with Ar gas in a Fishione model 1010. The sample was plasma cleaned for 4~min in a Fishione model 1020 in order to avoid carbon contamination.

The measurements were performed with a monochromated and probe-corrected FEI Titan G2 $60-300$~kV TEM. The microscope was operated at a high tension of 60~kV in order to increase the interaction cross-section, while also reducing any Cherenkov losses \cite{stoger-pollach_influence_2007, erni_impact_2008}. The final sample thickness of less than 30~nm further minimized such unwanted retardation loss effects efficiently. STEM was set up with a convergence angle of 66~mrad, and EELS was measured with a Gatan Quantum 965 GIF with a collection angle of 16.8~mrad. The signal was dispersed to 0.01~eV per channel, and the exposure time of each spectrum in the Spectrum Image was set to right below the overexposure limit of the CCD. By using the monocromator, the energy resolution was 0.12~eV by measuring the full width at half maximum (FWHM) of the ZLP. After measurements, the data sets were corrected for dark current and energy calibrated by aligning the maximum of the ZLP of each spectrum to the same channel.

%%%%%%%%%%

\section{Results and Discussion}

%%%%%

\subsection{Manual fitting of background and edge onset}

\begin{figure}[tb]
\centering
\includegraphics[width=\linewidth,keepaspectratio]{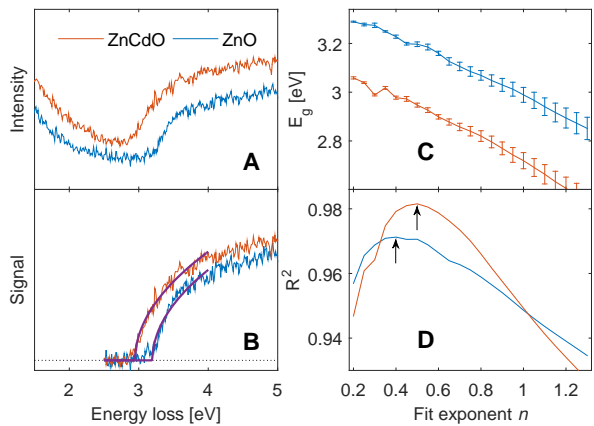}
\caption{\label{fig:fitfunc}(A) Low-loss EELS in ZnO and ZnCdO, 10$\times$ binned, and (B) background-subtracted spectra with least squares onset fit, $n$=0.5. (C) Fitted band gap as a function of curve exponent $n$ from least squares fit, and (D) corresponding goodness-of-fit parameter \Rsq. The arrows indicate the maximum. }
\end{figure}

Our starting points are two low-noise spectra obtained from ZnO and ZnCdO. From the 10$\times$ binned data in Fig.~\ref{fig:fitfunc}A it can be seen that the edge onsets are located around 3.2~eV in ZnO and 2.9~eV in ZnCdO. A successful removal of the background can be achieved with a good model, fitted to an energy range close to, but not overlapping with, the edge onset. We chose a decaying power-law model, and by manual inspection of the residual intensity (Fig.~\ref{fig:fitfunc}B), suitable background fit ranges were found at $2.5-2.8$~eV and $2.0-2.3$~eV for ZnO and ZnCdO, respectively. If regions closer to or further from the edges were chosen, we observe that the background is often over- or underestimated. This underlines the importance of the choice of fitting range for the background subtraction: the background in a Spectrum Image where the onset varies should be fitted relative to the edge and not in a fixed range of energy loss. 

To identify the band gap, we build on the work of Rafferty and Brown\cite{rafferty_direct_1998}. Based on an idealized band structure consisting of two parabolic bands describing the valence and conduction states, the observed energy loss intensity is described as
\begin{equation}
I(E) = c (E-E_\text{g})^n \; \text{ for }E \geq E_\text{g}.
\label{eq:exp}
\end{equation}
Here $E$ is the energy loss, \Eg is the band gap, and $c$ is a constant. The exponent $n$ is ideally $1/2$ for a direct band gap and $3/2$ for an indirect band gap. The model describes the ideal EELS edge in a short energy range above the onset, where the parabolic approximation holds. Below the onset, the ideal intensity is zero. In the present work, we limit ourselves to direct band gaps. Indirect band gap materials usually have a gradual onset of energy loss (Eqn.~\eqref{eq:exp}) which is easily masked by noise, background, or Cherenkov losses. Modified experimental setups are often needed to study these, such as allowing only specific momentum transfers to contribute to the spectrum\cite{gu_band-gap_2007}.

To test the suitability of Eqn.~\eqref{eq:exp}, we first manually identify an acceptable edge fitting range for both materials as $2.5-4.0$~eV. As edge onset is a one-sided fitting, any remaining intensity below the onset will shift the result down in energy. It is therefore more likely to find a lower than a higher onset of the edge. Hence, we seek both a high goodness-of-fit and a high onset when performing the fitting. By varying the exponent $n$, we find the band onsets as shown in Fig.~\ref{fig:fitfunc}C, and the corresponding goodness-of-fits (\Rsq) shown in Fig.~\ref{fig:fitfunc}D. The maximum of \Rsq is found at $n=0.45$ in ZnO and $n=0.55$ in ZnCdO, and this is seen to vary with the choice of onset fit range. As both materials are direct band gap semiconductors, and the model is limited to a small energy range above the onset, we assume that it is sufficient to use $n=0.5$ in both cases.

%%%%%

\subsection{Automated approach for edge onset extraction}

\begin{figure}[tb]
\includegraphics[width=\linewidth,keepaspectratio]{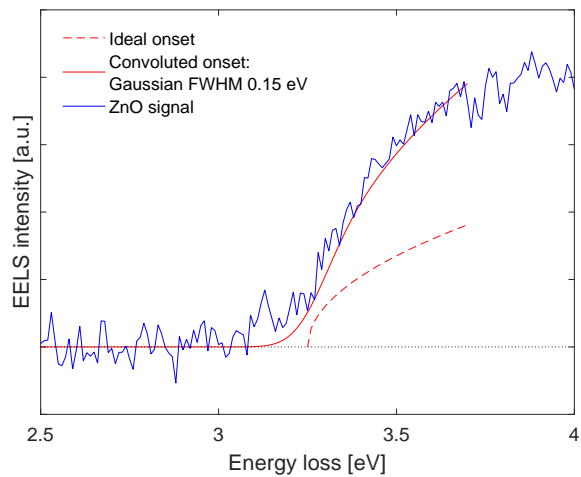}%
\caption{\label{fig:eres}The ideal edge (arbitrarily scaled) with a simulated energy broadening, obtained by Gaussian convolution, compared with the ZnO spectrum.}
\end{figure}

\begin{figure*}[tb]
\includegraphics[width=.33\linewidth,keepaspectratio]{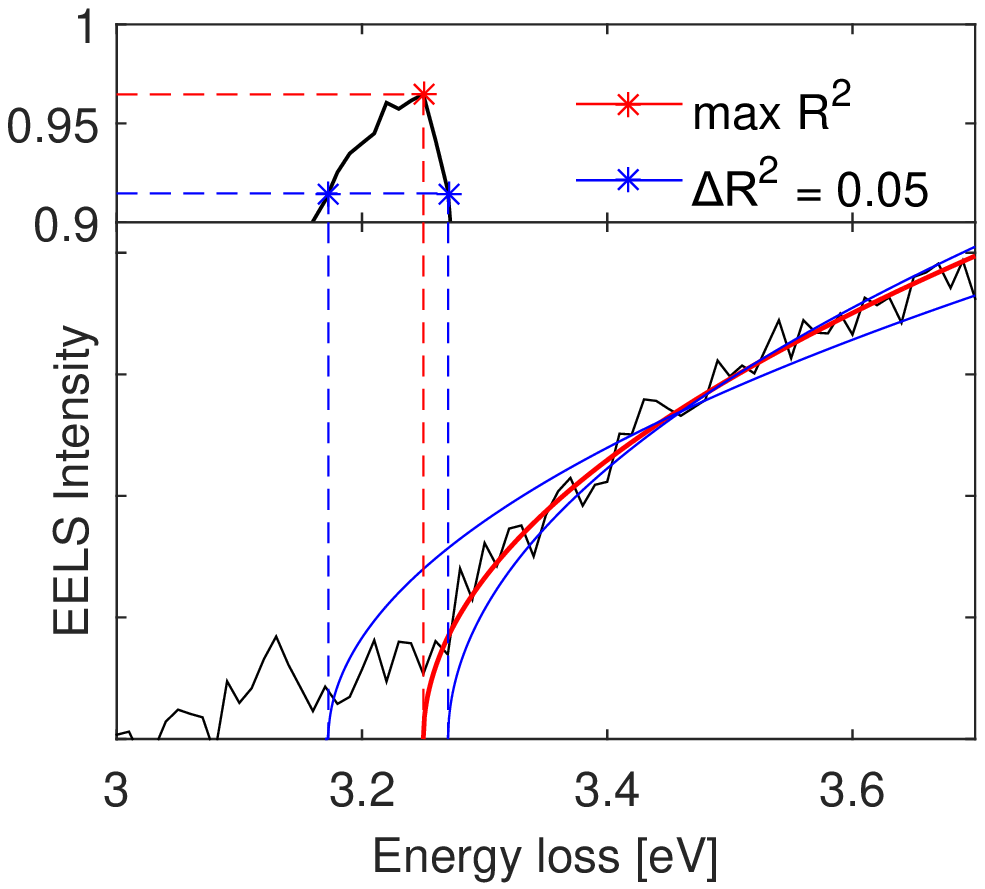}%
\includegraphics[width=.33\linewidth,keepaspectratio]{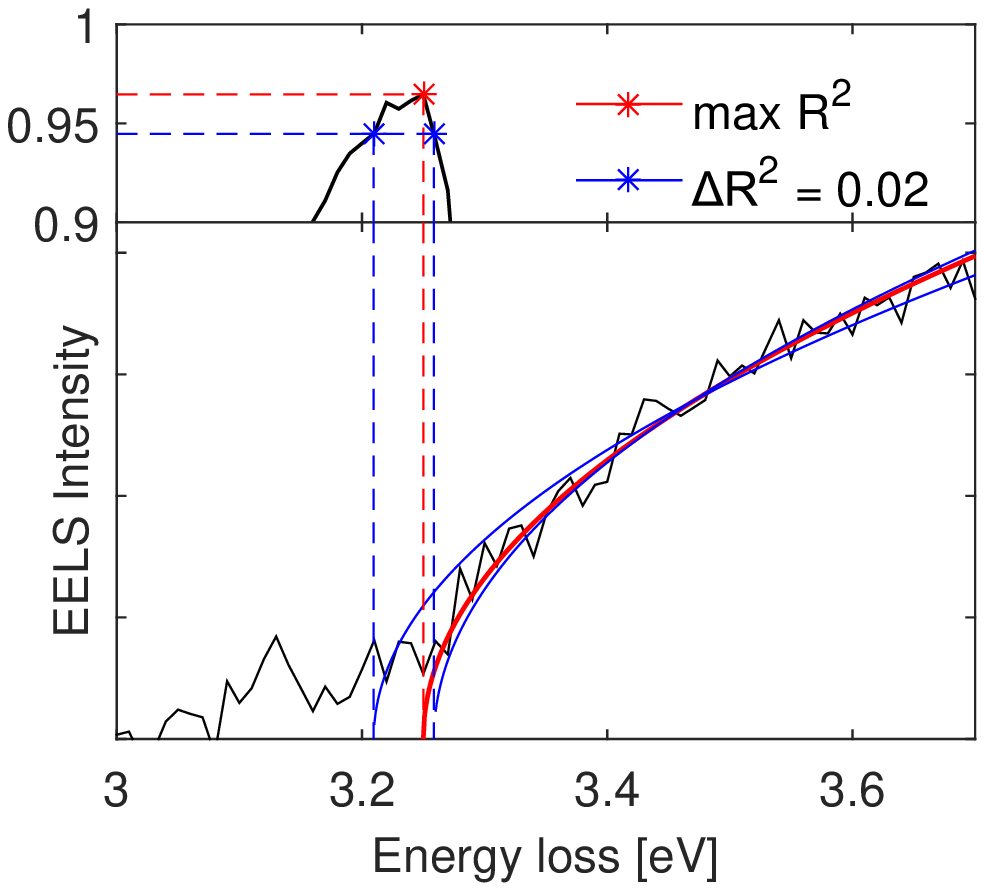}%
\includegraphics[width=.33\linewidth,keepaspectratio]{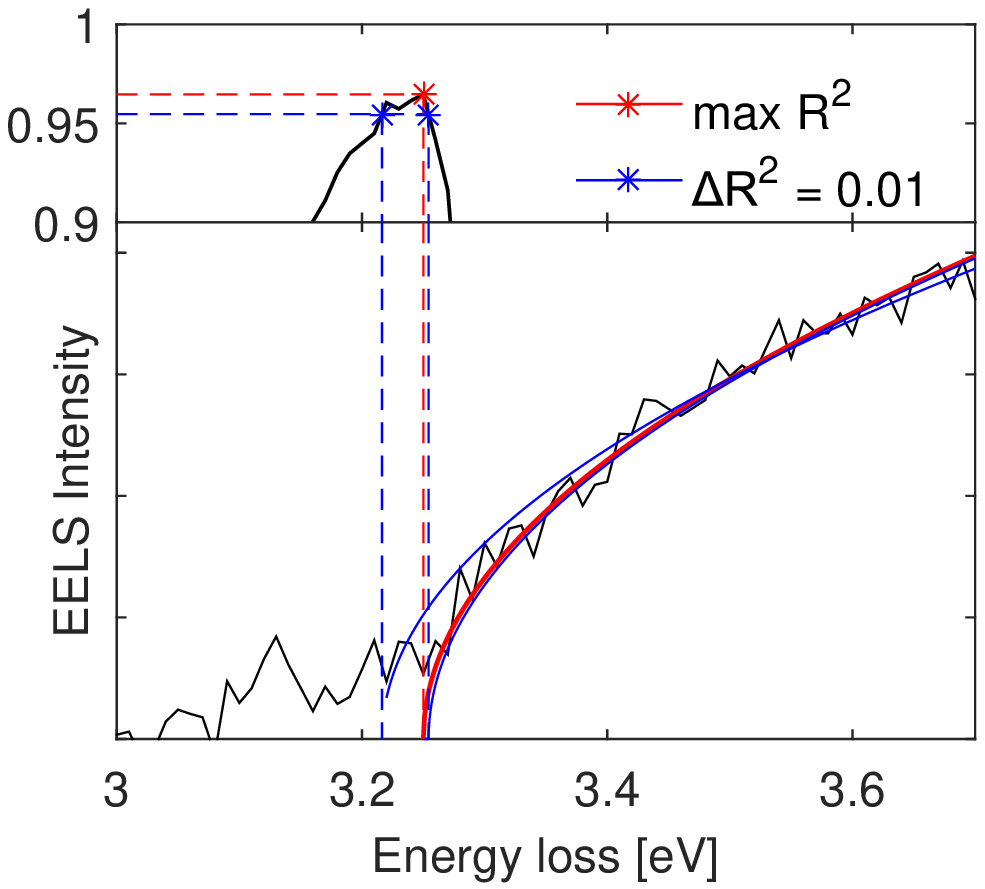}%
\caption{\label{fig:errorbars}Curve fitting of ZnO EELS spectrum, where the band gap is the onset of the fit which maximizes \Rsq. Error bars can be estimated from the reduction in \Rsq, here showing the fitted lines from reductions of 0.05, 0.02 and 0.01 in \Rsq value. }
\end{figure*}

\begin{figure}[tb]
\includegraphics[width=.47\linewidth,keepaspectratio]{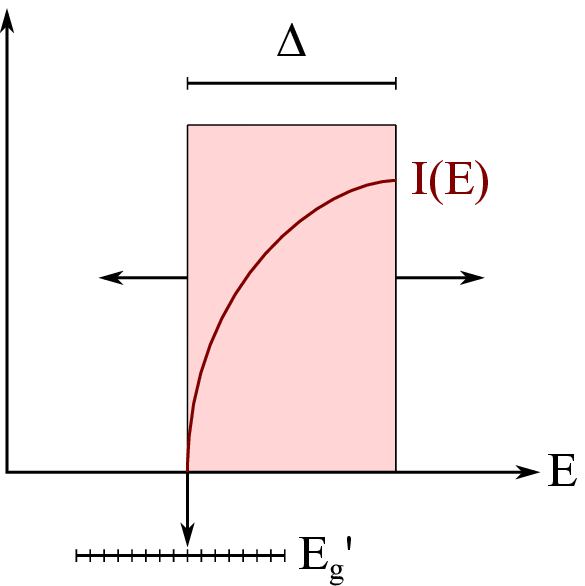}%
\hfill%
\includegraphics[width=.47\linewidth,keepaspectratio]{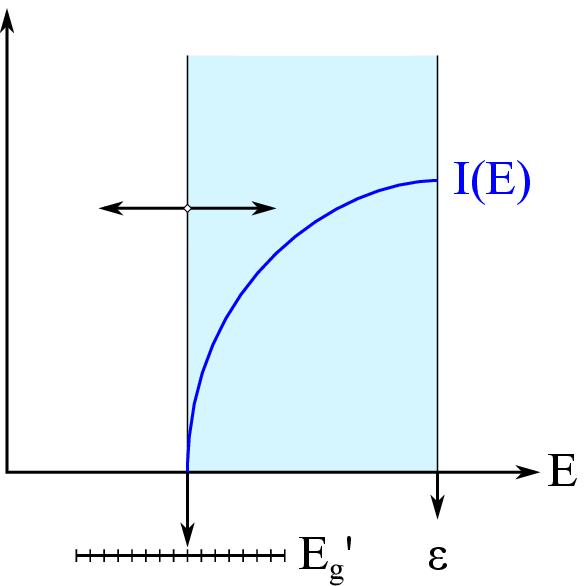}
\caption{\label{fig:fitfigs}Illustrations of the Sliding interval fit method with parameter $\Delta$ (left) and the Fixed endpoint method with parameter $\varepsilon$ (right). \Egp illustrates the range of test values for edge onset and $f(E)$ is the fitted model. }
\end{figure}

From Eqn.~\eqref{eq:exp} the intensity below the onset is ideally zero, however, this is rarely the case for experimentally obtained spectra. In addition to noise and residual signal after background subtraction, the energy resolution of the experiment is assumed to introduce a spectral broadening which affects the edge. The simulated impact of spectral broadening is shown in Fig.~\ref{fig:eres}, where the ideal onset convoluted with a Gaussian (FWHM is set to 0.15 eV) can be compared with the ZnO spectrum. Here the energy broadening creates a tail to the ideal spectrum, which leads to intensity below the onset. When making a curve fit to this edge, this intensity can easily give a lower onset energy, which would be interpreted as a lower band gap value. These experimetal factors can normally be reduced by deconvolution procedures, where deconvolution for background removal, noise reduction and/or energy sharpening are possible approaches, however, deconvolution is known to work poorly on noisy spectra\cite{wang_fourier-ratio_2009, kimoto_advantages_2005}. 

A different approach is to exclude this intensity from the edge fitting region. More specifically, the fitting range should depend on the onset. This is difficult to implement when using methods where the onset is an output of the fitting itself, such as a regular least squares fit of Eqn.~\eqref{eq:exp}. For analysis of only a handful of spectra, manual inspection of the fitting range can correct this error, but for large Spectrum Images with a variable onset this is not a feasible approach. We therefore choose an alternative approach, where a range of possible onsets are tested and evaluated according to their goodness of fit. A reasonable range of test values can usually be identified, either from prior knowledge of the material, or by visual inspection of the raw data. For each energy point in this range the appropriate energy-loss model of Eqn.~\eqref{eq:exp} is fitted, and a goodness of fit is calculated. The onset is then identified as the test value giving the best goodness of fit. In this approach, the lower limit of each fitting is the test value itself, and only intensity above this value is used in the fitting and evaluation. Any tail from energy broadening, noise or spurious signal below the test value is ignored, and a more robust identification of the onset can be made.

As a measure of the goodness of fit, we use the coefficient of determination \Rsq, and the onset \Eg is found as the test value \Egp which maximizes \Rsq. To indicate the uncertainty in the onset determination, we define error bars giving a certain reduction of \Rsq. Fig.~\ref{fig:errorbars} shows an example of optimal fitting of ZnO together with fittings performed at upper and lower test values corresponding to the reduction in \Rsq of 0.05, 0.02, and 0.01. These plots are visually inspected to identify a level of uncertainty that the user finds suitable, based on the level of noise, accuracy of the fit, and the level of confidence required for the current application. In large data sets with varying levels of noise different levels of \Rsq reduction could be the optimal choice, however, as the level of noise is similar throughout the current data set, studying only a few spectra is sufficient. For this 10$\times$ binned ZnO spectrum, an \Rsq reduction of 0.05 results in excessive deviations from the optimal fit, resulting in error bars that we judge as unreasonably large. In comparison, an \Rsq reduction of 0.01 leads to the upper fit being closer to the optimal one than one channel, which we consider unphysical. A reduction of 0.02 seems to be a reasonable choice, and similar conclusions were drawn from investigations of the 10x binned ZnCdO spectrum.

\begin{figure*}[tb]
\begin{minipage}[b]{.49\linewidth}
\includegraphics[width=\linewidth,keepaspectratio]{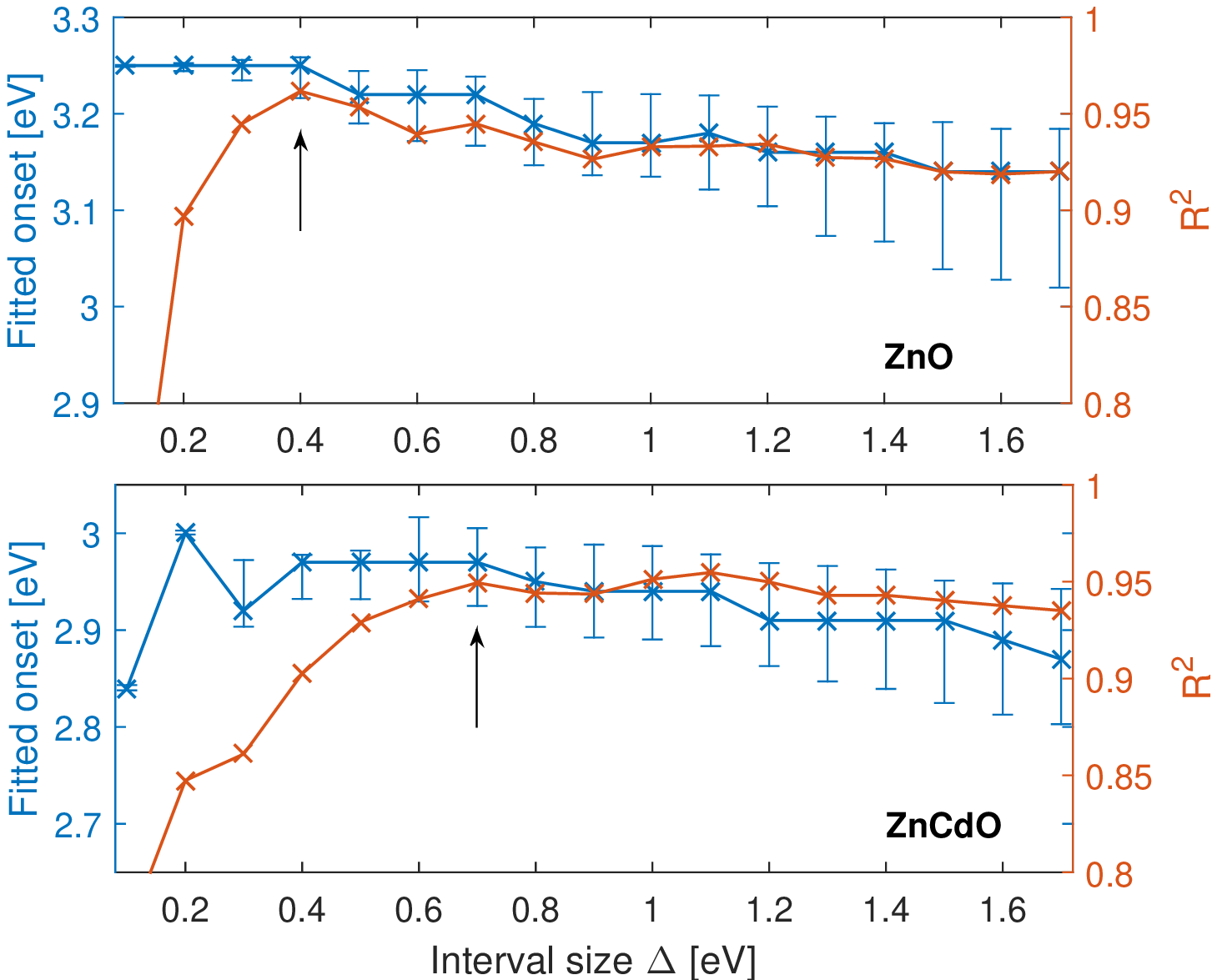}
\caption{\label{fig:slide}Investigation to find the optimal fit parameter \del in the Sliding interval method, evaluated on ZnO (top) and ZnCdO (bottom) edges. The onset is the tested value resulting in maximal \Rsq, with the errorbars as the interpolated energies of a 0.02 decrease in \Rsq.}
\end{minipage}%
\hfill
\begin{minipage}[b]{.49\linewidth}
\includegraphics[width=\linewidth,keepaspectratio]{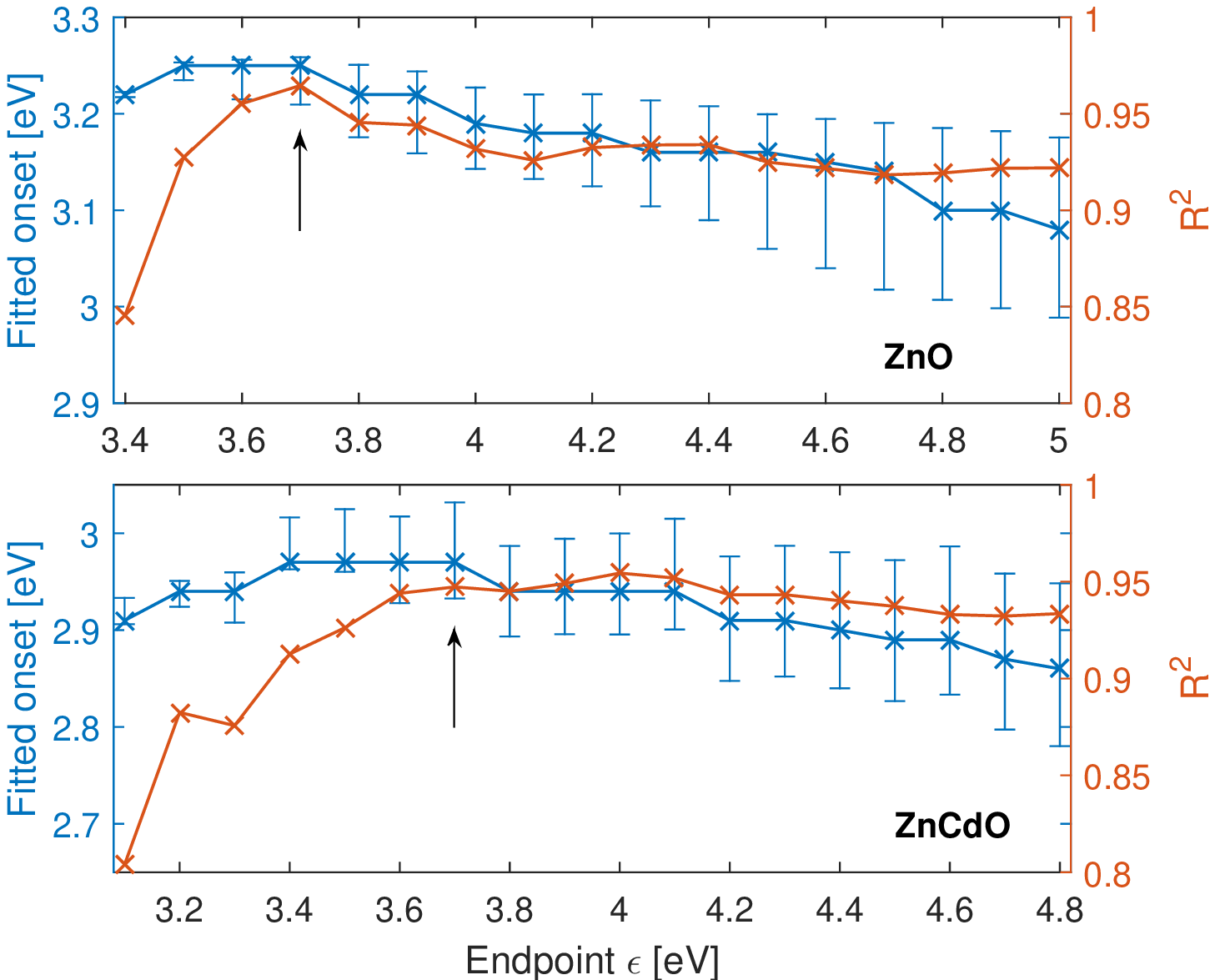}
\caption{\label{fig:endpt}Optimization of fit parameter \eps in the Fixed endpoint method, evaluated on ZnO (top) and ZnCdO (bottom). The fitted onset of the edge, \Rsq and interpolated energies from a 0.02 decrease of \Rsq is plotted for each tested endpoint. Note the shifted \eps-axis.}
\end{minipage}
\end{figure*}

As mentioned previously, Eqn.~\eqref{eq:exp} is based on models where the only transitions that occur are from a single parabolic valence band to a single parabolic conduction band. This may be a good assumption for energies close to the band gap value, but for higher energies transitions between additional valence and conduction states will occur. It can therefore not be expected that the shape of the energy loss will be faithfully reproduced by this model over a large energy range. The choice of upper bound of the fitting procedure therefore becomes crucial. Again, this can be handled manually if only a handful of spectra are to be analysed, or if there is very little variation in electronic structure in the data set to be analysed. However, for any automatic analysis of large inhomogeneous data sets this is problematic. 

We have therefore implemented two different methods that can be used in automated band gap extraction. First, based on the work by Rafferty and Brown\cite{rafferty_direct_1998}, an approach we refer to as the `Sliding interval' method uses an interval with a fixed energy range relative to the test value \Egp. Specifically, for each test value, a range from test value \Egp up to the point \Egp + \del is used in the fitting, where \del is a constant. As the whole series of values \Egp is tested, the fitting range then `slides' across the spectrum. Second, a `Fixed endpoint' method is also implemented. Here the end of the fit range is a fixed energy \eps, and as the whole series of \Egp is tested, the size of the fitting window varies. The two approaches are illustrated in Fig.~\ref{fig:fitfigs}.

%%%%%

\subsection{Fit parameter determination}

As the overall goal is to find a method which can be applied to a full Spectrum Image, it is important that the same parameters for the fitting are valid for every single spectrum. This means that the range of test values must contain all possible edge onsets within the Spectrum Image, and also that the same end-of-fit parameter \del or \eps must be applicable to all spectra. In fact, the end-of-fit parameter should be optimal in each spectrum that is to be fitted to get the most accurate result. To achieve this, the extreme types of spectra in the Spectrum Image should be inspected to find the range where the end-of-fit parameter is optimal, and the overlap of optimal parameter determines which parameter should be used for the full Spectrum Image analysis.

The shape of the edges and the overlap of optimal fitting parameter determine which of the two methods is best for a particular data set. The Sliding interval method is expected to be the best choice if the edges in the Spectrum Image have similar slopes and only a shift in onset, however, if there is a large variation in the edge steepness, a fixed width $\Delta$ may be difficult to define optimally for all spectra. On the other hand, the Fixed endpoint method is expected to work well if the edge becomes less sharp with decreasing onset. However, if the lowest onset require a narrower fit range than one describing the high onset spectrum, problems should be expected with this method. 

To investigate and illustrate the performance of the two end point methods, we use the 10$\times$ binned and manual background subtracted SI shown in Fig.~\ref{fig:fitfunc}B. In the Sliding interval method a range of different \del values are tested while recording the extracted onset \Eg, the corresponding \Rsq, and the interpolated onset error of a 0.02 decrease of \Rsq. The result is shown in Fig.~\ref{fig:slide}. The optimal parameter \del is found through a combination of high onset value, small errors, and high \Rsq, and are found to be $\Delta_\text{ZnO}=0.4$~eV and $\Delta_\text{ZnCdO}=0.7$~eV. With the optimal conditions of ZnO, the fit of the ZnCdO gives an error of 0.01~eV, equal to one channel. 

For the Fixed endpoint method a range of different \eps values are tested. The result is shown in Fig.~\ref{fig:endpt}, showing the optimal parameters are $\varepsilon_\text{ZnO}=3.7$~eV and $\varepsilon_\text{ZnCdO}=4.0$~eV. While the results for the ZnO layer depends heavily on the choice of \eps, the output for ZnCdO is less dependent on the choice of endpoint. The same parameter $\varepsilon_\text{ZnO}$ can be used for both layers without introducing systematic errors. This confirms that the Fixed endpoint method is the best method for the system, when optimized for ZnO.

%%%%%

\subsection{Dynamic background subtraction}

\begin{figure}[tb]
\includegraphics[width=\linewidth,keepaspectratio]{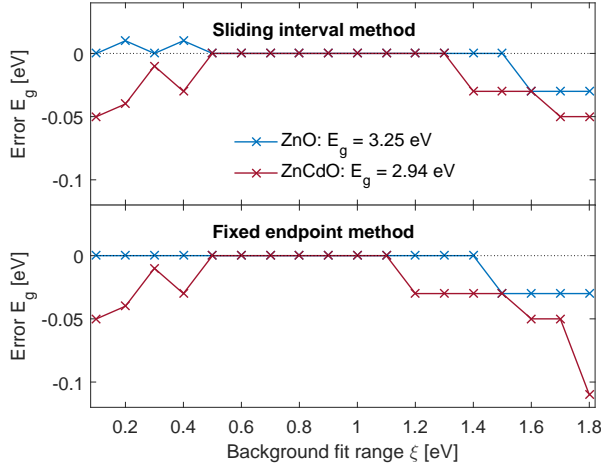}
\caption{\label{fig:bg1}Fitted onset error as a function of the range of dynamic background subtraction $\xi$, using the Sliding interval method (top) and Fixed endpoint method (bottom).}
\end{figure}

One of the main challenges of extracting accurate band gaps from experimental data is the removal of the background formed by the tail of the ZLP. Although this tail is greatly supressed in monochromated instruments, and generally has a low intensity in the relevant energy ranges for wide band gap semiconductors, the quality of the background subtraction can still have a significant impact on the extracted values. 

Several approaches for background removal have been described in literature, often resulting in slightly different results of the band gap fit\cite{gu_band-gap_2007, stoger-pollach_optical_2008}. As mentioned previously, deconvolution methods are not ideal where the noise level is high, and curve fitting is assumed to be a better approach for background removal\cite{zhan_nanoscale_2017_fix}. Such curve fitting methods include fitting using a reference ZLP measured in vacuum, or mirroring the left, negative side of the ZLP, but a simpler approach is found by assuming the background follows a power law function. We have previously found this to be a suitable approach for the wide band gap semiconductors considered in this work, any may even be preferable to the more advanced methods. However, the extracted onset is often still found to depend heavily on the fitting range used for the background subtraction, in particular for samples and systems where the position of the edge onset varies\cite{zhan_nanoscale_2017_fix}. For such systems, a single choice of background fitting range may not be suitable. 

In the present approach, the edge onset value is identified from a range of pre-determined test values, and is not an output of the fitting itself. As a consequence, we can use these test values as input for a background subtraction method that can handle onset variations dynamically. Specifically, we have implemented a method where each test value of the onset \Egp is used to define background fitting windows \Egp - \x to \Egp (or to \Egp-$\delta$, where $\delta$ is a small, fixed range between background fit range and onset) for each pixel. The corresponding background subtracted signals are then used in one of the fitting procedures described above to identify the test value that gives the highest \Rsq. In this way, an optimal background subtraction can be performed for each pixel, and only the parameter \x must be set, corresponding to the size of the background fitting range.

This method is tested on the same ZnO/ZnCdO system as previously in order to find optimal values of the background fitting width \x. The results are shown in Fig.~\ref{fig:bg1}, where the dynamic background subtraction is applied in combination with the Sliding interval method, and the Fixed endpoint method, respectively. In both approaches there is a significant range of overlap where the onset error is zero, indicating where the optimal parameter of \x is. As the manual background subtraction range has been optimized in each of the spectra, and cannot be generalized to a full Spectrum Image without introducing errors, it can be seen that the dynamic background subtraction manages to capture the optimal background subtraction with a single parameter, thus performing better in a full Spectrum Image analysis.

%%%%%

\subsection{Handling of noise}
\label{sec}

\begin{figure}[tb]
\includegraphics[width=\linewidth,keepaspectratio]{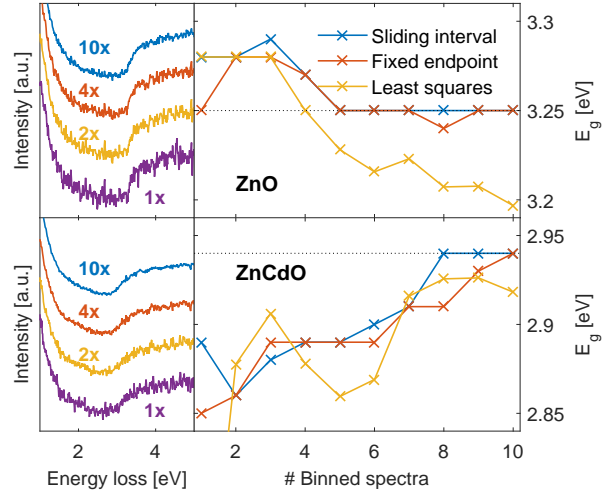}
\caption{\label{fig:spectra}Spectra of ZnO and ZnCdO with different binning (left), and fitted edge onset as a function of binning (right).}
\end{figure}

When applying the presented method to a full Spectrum Image, the level of noise may be important for the performance of the method. The above analysis was performed with a relatively high (10$\times$) binning, which may not be suitable when seeking a high spatial resolution mapping. However, lower binning increase the level of noise, as can be seen from the differently binned spectra in Fig.~\ref{fig:spectra} (left). Here the unbinned spectrum has a high level of noise, which reduces the precision of an extracted onset. Fig.~\ref{fig:spectra} (right) shows the fitted onset as a function of binning in ZnO and ZnCdO, analyzed by the manually subtracted background with optimal fit parameters. Here it can be seen that a small random variation in fitted onset occurs due to random noise.
 
\begin{figure*}[t]
\includegraphics[width=\linewidth,keepaspectratio]{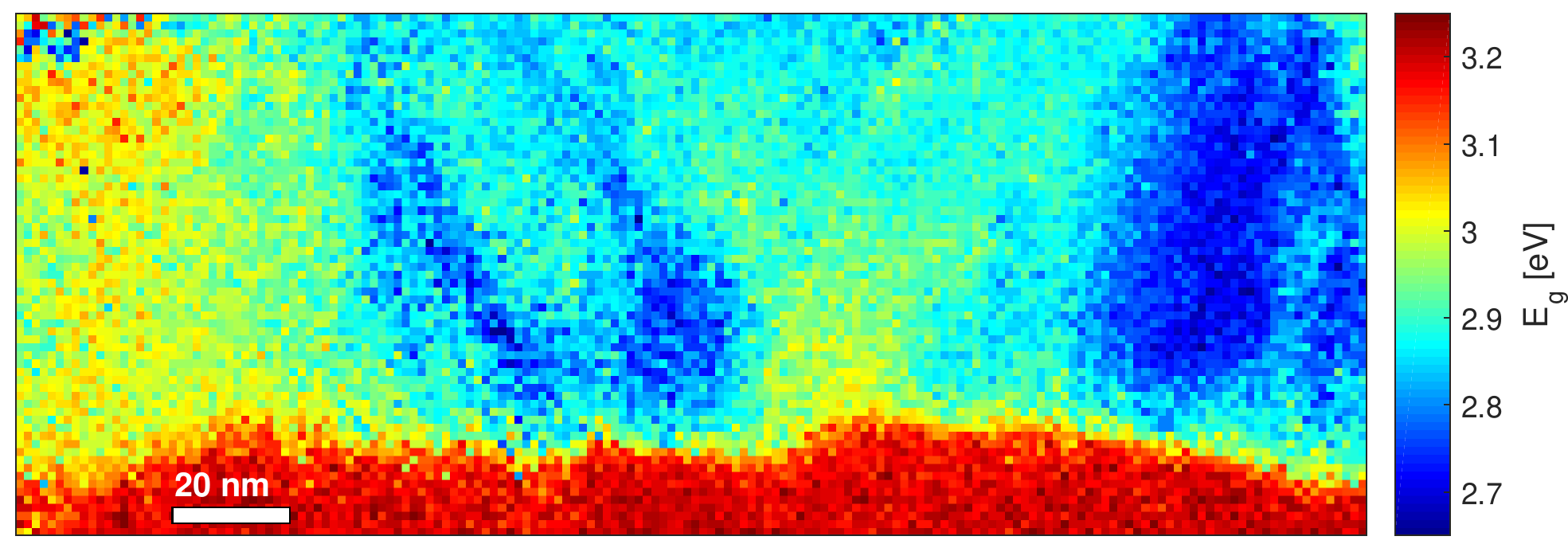}
\caption{\label{fig:map}Band gap mapping of ZnO/ZnCdO interface.}
\end{figure*}

The spectra analyzed in Fig.~\ref{fig:spectra} were manually background subtracted using the same fit range, and the band gaps were fitted with the Fixed endpoint and Sliding interval methods with the optimal parameters found above. The two methods produce similar onset values, which indicates that they are almost equally sensitive to the level of noise. In ZnO it can be seen that the error is within a few channels even at low binning, which can be explained by the ZnO edge being relatively sharp and well defined, so that both fitting methods are able to extract the onset with relatively high precision. In ZnCdO the extracted onset is reduced at lower binnings, which is assigned to a high level of noise in combination with a less sharp edge, giving a lower signal-to-noise ratio than in the ZnO spectra.

As a comparison, a regular least squares fit was also tested on the differently binned spectra. A fitting range of $2.0-4.0$~eV was used for the edge fitting of both spectra, and a function with zero below the onset was fitted with a least squares method. As can be seen from Fig.~\ref{fig:spectra} the deviations of the two methods described here are smaller than from the least squares method. This indicates that the Sliding interval and Fixed endpoint methods have higher precision than a regular least squares fit, and can be explained by a reduced impact of the noise level below and around the onset when using test values for the onset.

%%%%%%%%%%

\section{Spectrum Image fitting}

Throughout the analysis we have developed band gap extraction methods which are customized for ZnO and ZnCdO systems. A large Spectrum Image acquired over the ZnO/ZnCdO interface has been used to test the performance of the method, and was imported into Matlab. The energy scale of the Spectrum Image was calibrated by aligning all ZLP's to the maximum value at 0~eV, and in order to reduce the noise level a 2$\times$2 binning was applied. According to the results in section~\ref{sec}, binning 4 spectra together would result in an error less than 0.04~eV in both regions. Also, this reduced the total number of spectra in the full data set to 10~920. The onset error bars were calculated from a 0.05 decrease in \Rsq, and the fitting was set up with test values in the range \Egp = [2.45, 3.35]~eV, which gives 91 test values for each spectrum. The Fixed endpoint method was used with \eps = 3.70~eV, and the dynamic background subtraction was applied in the range \x = 0.70~eV below the test values (a small region $\delta$ = 0.10~eV was discarded between the background and onset). 

The fitting was performed with Matlab 2016b with 16 workers, running under Windows Server 2016 on an Intel Xeon E5-4640 processor with 164 GB of RAM. The full analysis was completed in 2 hours and 48 minutes, effectively giving 14.8 s per spectrum per worker, depending on the number of test values. The resulting band gap map is shown in Fig.~\ref{fig:map}, and the corresponding error mappings are shown in Supplementary Information. The bottom shows the ZnO layer, where the average onset is found at 3.20 eV, and where the standard deviation of 0.02~eV indicates the precision of the extracted onsets. The experimental band gap value is somewhat lower than usually found in ZnO, though detailed interpretation together with previously published Cathodoluminescence spectroscopy measurements\cite{zhan_nanoscale_2017_fix} confirms the accuracy of the measurement, which suggests that the mapping outcome is convincing. Furthermore, local band gap variations can be identified within the ZnCdO layer, caused by uneven Cd content. This is supported by local variations in Cd concentration found by energy-dispersive X-ray spectroscopy (EDS) mapping (not shown), and has been shown to have little correlation with the sample thickness (details in Supplementary Information). Then, by this method, we are able to directly map the effect of composition variation in band gaps.

%%%%%%%%%%

\section{Conclusions}

In this work, new methods for dynamic extraction of optical band gaps from STEM-EELS Spectrum Images have been developed. Band gap extraction from EELS usually involve curve fitting to locate the edge onset, and in a Spectrum Image analysis where the onset may vary, this can introduce issues regarding the position and width of the fitting ranges, handling of certain spectral features, and effects of a high level of noise. The methods described in this work rely on input parameters rather than fixed fitting ranges, which enable the extraction of a large range of onset values. The implemented methods consist of an iterative procedure of checking possible values for the onset, which has the advantage of reducing the effect of experimental spectral broadening which can lower a fitted onset. In addition, a dynamic background subtraction method can be applied in combination with the onset fitting, thus reducing errors associated with insufficient background subtraction. The methods are shown to have high accuracy and precision when applied to a ZnO/ZnCdO system, where local band gap variations can be extracted. By this automatic method, large data sets can be analyzed, which opens up for exploring band gaps with a high spatial resolution.

%%%%%%%%%%

\section*{Acknowlegdements}

The Research Council of Norway is acknowledged for the support to the Norwegian Center for Transmission Electron Microscopy, NORTEM (project 197405/F50) and the Norwegian Micro- and Nano-Fabrication Facility, NorFab (project 197411/V30). Asbjørn Ulvestad is also acknowledged for useful discussions and comments.

%%%%%%%%%%

%\section*{Supplementary information}

%Supplementary material associated with this article can be found in the online version.

%%%%%%%%%%

%\section*{References}

\bibliography{bandgapprogram,bandgapprogram2}

\end{document}